\documentclass[a4paper,11pt]{article}
\usepackage{pos}
\pdfoutput=1
\usepackage{enumitem}

\title{Massive form factors at $\mathcal{O}(\alpha_s^3)$}

\author*[a]{Kay Schönwald}

\affiliation[a]{Institut für Theoretische Teilchenphysik, Karlsruhe Institute of Technology (KIT),\\
76128 Karlsruhe, Germany}

\emailAdd{kay.schoenwald@kit.edu}

\abstract{
We report on our recent calculation of massive quark form factors using 
a semi-numerical approach based on series expansions of the master integrals
around singular and regular kinematic points and numerical matching.
The methods allows to cover the whole kinematic range of negative and 
positive values of the virtuality $s$ with at least seven significant digits accuracy. 
}

\FullConference{
  Loops and Legs in Quantum Field Theory - LL2022,\\
  25-30 April, 2022\\
  Ettal, Germany 
}

\newcommand{\ii}{\text{i}}

\begin{document}
\maketitle

\section{Introduction}
Massive form factors are important objects in quantum field theory.
They constitute the virtual corrections to many observables and processes
like lepton pair production via the Drell-Yan process or the decay of 
the Higgs boson into heavy quarks.
The more and more precise measurements of these processes make the 
inclusion of higher order corrections for the theory predictions necessary.

Up to $\mathcal{O}(\alpha_s^2)$ the massive form factors are known in 
analytic form, see Refs.~\cite{Hoang:1997ca,Mastrolia:2003yz,Bonciani:2003ai,Bernreuther:2004ih,Bernreuther:2004th,Bernreuther:2005rw,Bernreuther:2005gw}, 
and even higher orders in the dimensional regulator $\epsilon$ haven been considered in 
Refs.~\cite{Gluza:2009yy,Henn:2016tyf,Ahmed:2017gyt,Ablinger:2017hst,Lee:2018nxa}.
At $\mathcal{O}(\alpha_s^3)$ only partial results are available.
At this order the form factors have been considered in the large-$N_C$ limit 
(with $N_C$ the number of colors) in 
Refs.~\cite{Henn:2016tyf,Lee:2018rgs,Ablinger:2018yae,Ablinger:2018zwz},
the light fermion contributions were calculated in Ref.~\cite{Lee:2018nxa}
and in Ref.~\cite{Blumlein:2019oas} all non-singlet contributions involving a closed
heavy quark loop have been considered.

In these proceedings we report on our recent calculation of the massive form factors 
at $\mathcal{O}(\alpha_s^3)$ in Refs.~\cite{Fael:2022rgm,Fael:2022miw}, 
where we employed a semi-numerical method involving series expansions and numerical 
matching between them.
In Section~\ref{sec:2} we will summarize technical details, while in Section~\ref{sec:3} 
we show some results.
In Section~\ref{sec:4} we conclude and give an outlook. 

\section{Massive form factors}
\label{sec:2}
\begin{figure}
    \centering
    \includegraphics[width=0.32\textwidth]{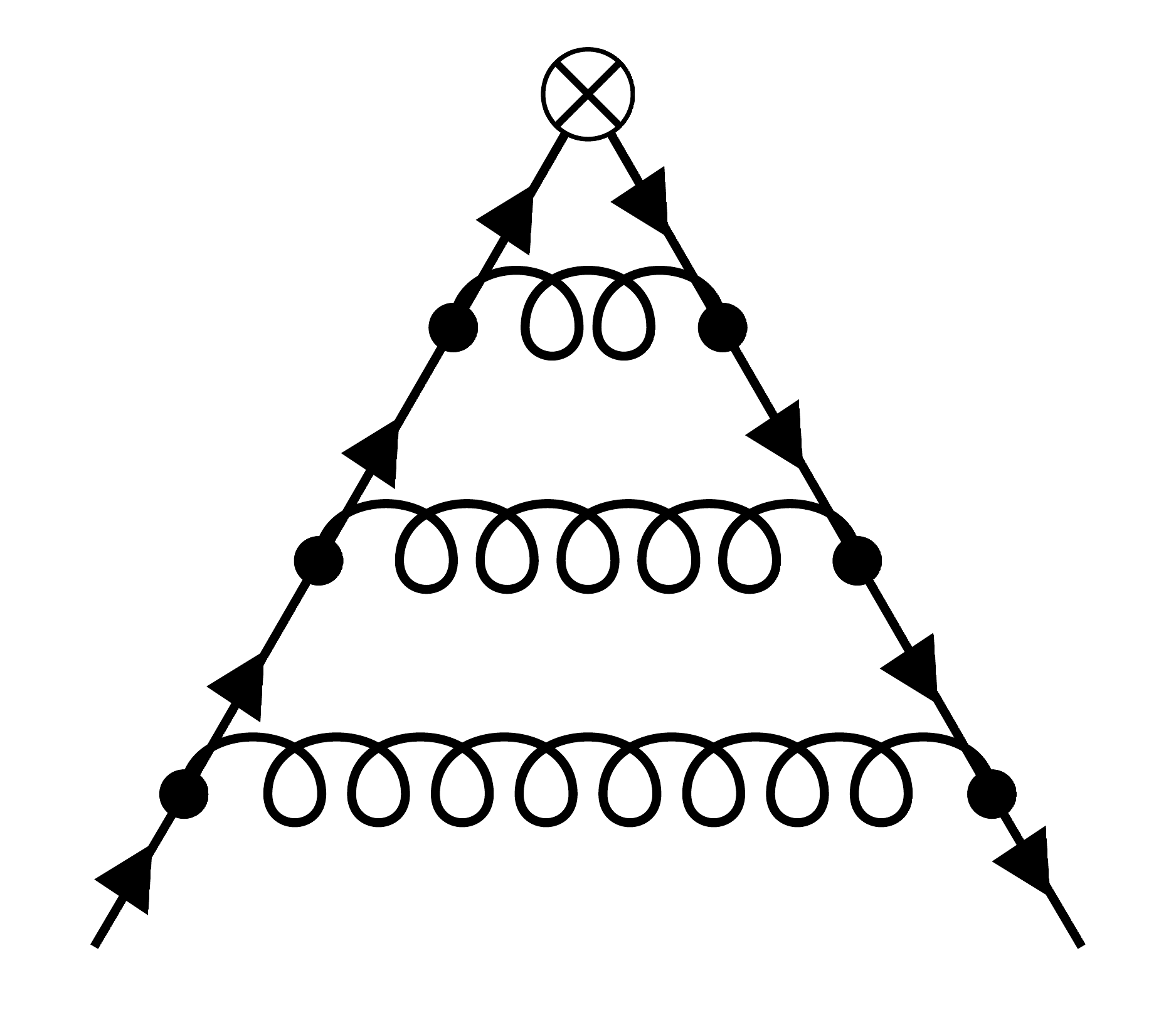}
    \includegraphics[width=0.32\textwidth]{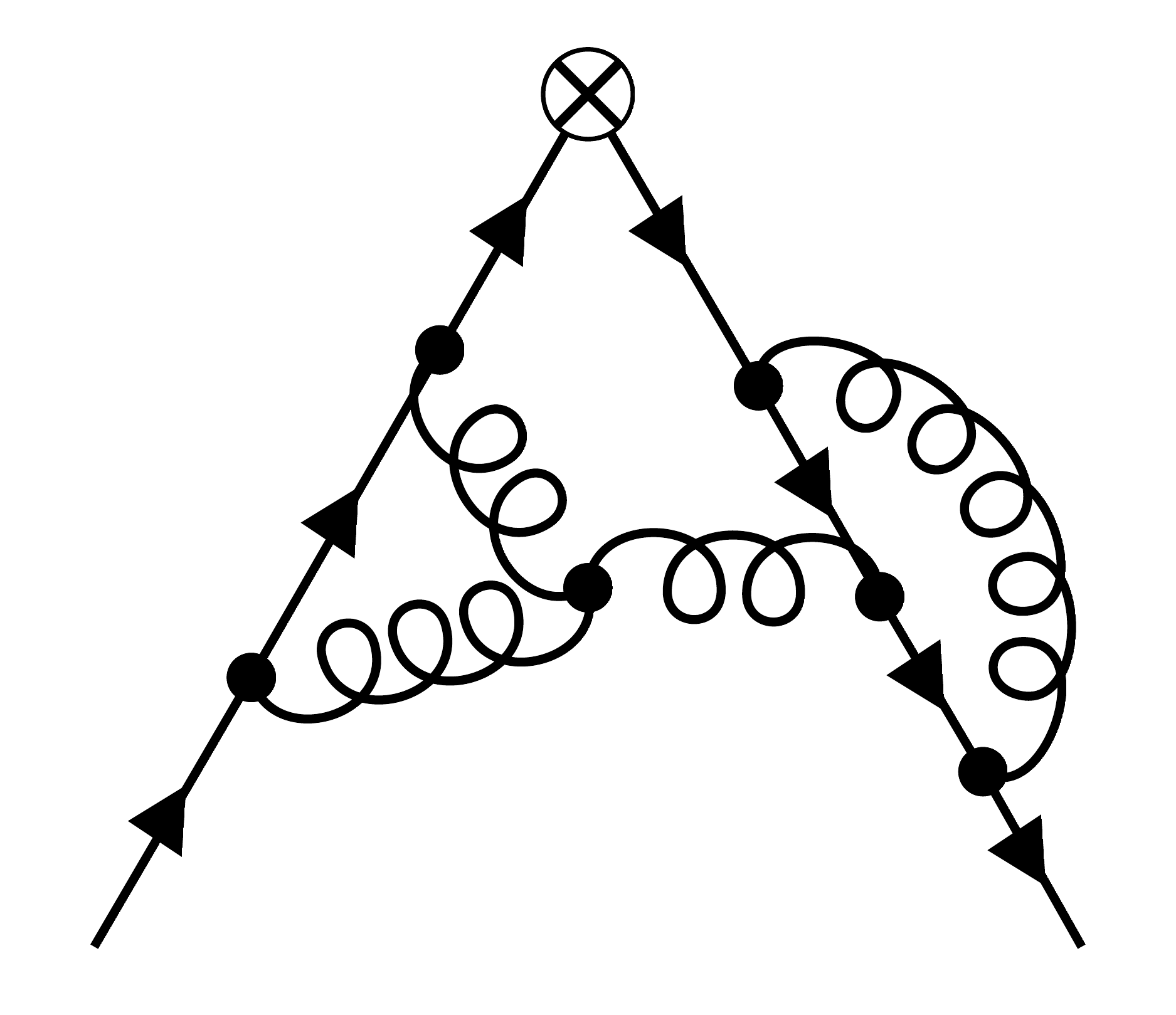}
    \includegraphics[width=0.32\textwidth]{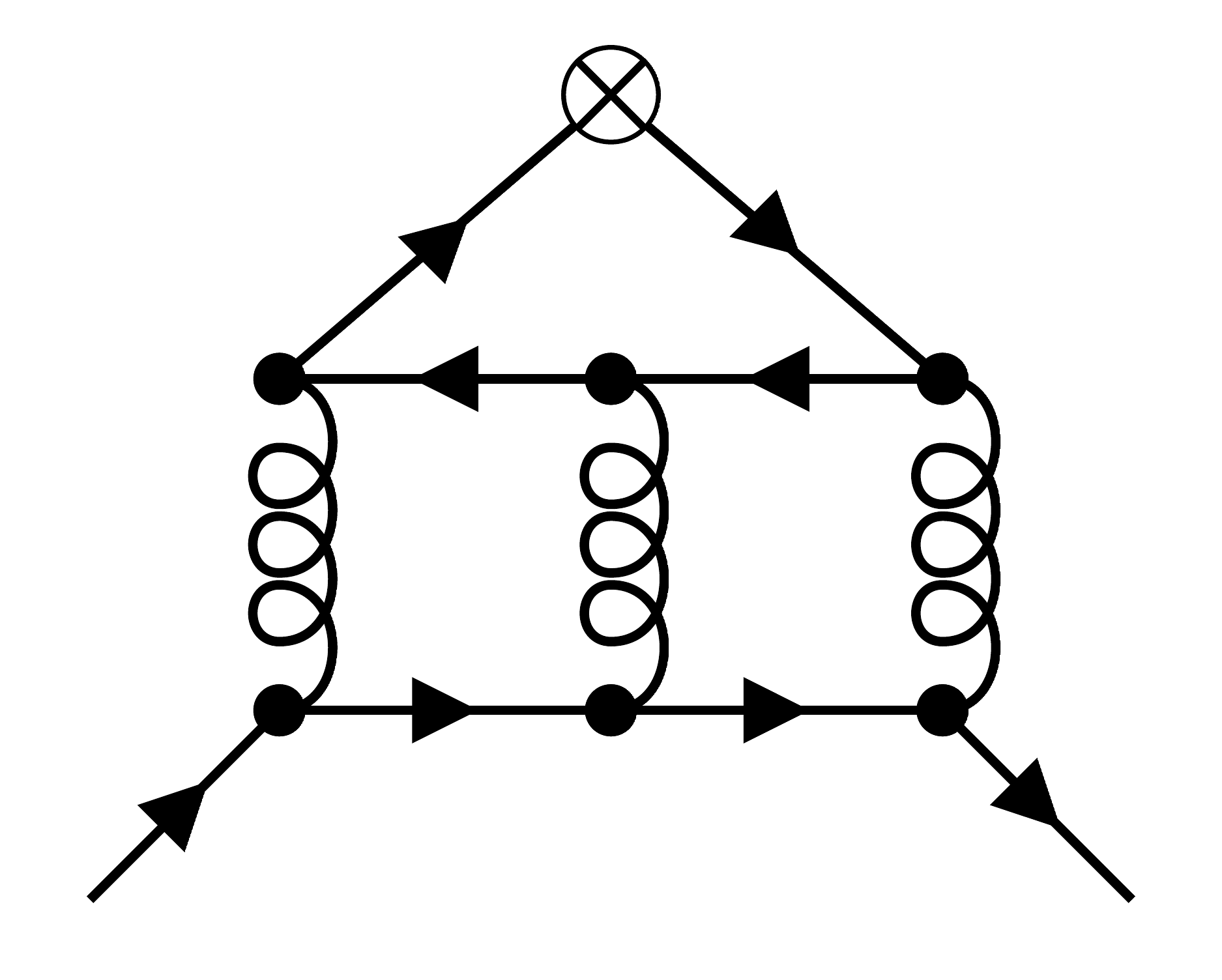}
    \caption{Sample Feynman diagrams contributing to the massive quark form 
    factors. Solid lines represent massive quark, while curly lines represent gluons. The vertex with a cross refers to the coupling to one of the external currents. The diagram on the right is part of the singlet contributions.}
    \label{fig::Graphs}
\end{figure}

To compute the massive form factors we consider the interaction of a 
massive quark with a vector, axial-vector, scalar or pseudo-scalar current,
which are given by:
\begin{eqnarray}
    j_\mu^v &=& \bar{\psi}\gamma_\mu\psi\,,\nonumber\\
    j_\mu^a &=& \bar{\psi}\gamma_\mu\gamma_5\psi\,,\nonumber\\
    j^s &=& m \,\bar{\psi}\psi\,,\nonumber\\
    j^p &=& \ii m \,\bar{\psi}\gamma_5\psi\,.
    \label{eq::currents}
\end{eqnarray}
The vertex function can then be expressed through six scalar functions by 
\begin{eqnarray}
    \Gamma_\mu^v(q_1,q_2) &=&
    F_1^v(q^2)\gamma_\mu - \frac{\ii}{2m}F_2^v(q^2) \sigma_{\mu\nu}q^\nu
    \,, \nonumber\\
    \Gamma_\mu^a(q_1,q_2) &=&
    F_1^a(q^2)\gamma_\mu\gamma_5 {- \frac{1}{2m}F_2^a(q^2) q_\mu }\gamma_5
    \,, \nonumber\\
    \Gamma^s(q_1,q_2) &=& {m} F^s(q^2)
    \,, \nonumber\\
    \Gamma^p(q_1,q_2) &=& {\ii m} F^p(q^2) {\gamma_5}
    \,.
    \label{eq::Gamma}
  \end{eqnarray}
Here the momentum $q_1$ ($q_2$) is incoming (outgoing), on-shell ($q_1^2=q_2^2=m^2$)
and $q=q_1-q_2$ is the outgoing momentum at the current $j^\delta$ with $q^2=s$.
The form factors have an expansion in the strong coupling constant $\alpha_s$
\begin{align*}
    F_{i}^{j} = \sum\limits_l \left( \frac{\alpha_s}{\pi} \right)^l F_{i}^{j,(l)} ~.
\end{align*}
We divide the form factor into non-singlet and singlet contributions, where the 
current couples to the heavy external quark line or an internal heavy quark loop,
respectively. 
Some sample Feynman diagrams contributing to the form factors can be found in 
Fig.~\ref{fig::Graphs}.

The calculation of the form factors proceeds in the following way:
        We generate the diagrams with \texttt{QGRAF} \cite{Nogueira:1991ex} and use \texttt{q2e} \cite{Harlander:1998cmq,Seidensticker:1999bb}
    to transform the output to \texttt{FORM} \cite{Ruijl:2017dtg} input, where Dirac-, Lorenz- and color-algebra 
    (with \texttt{color} \cite{vanRitbergen:1998pn}) is performed. 
    The diagrams are mapped to predefined topologies using \texttt{exp} \cite{Harlander:1998cmq,Seidensticker:1999bb}.
       The scalar integrals are reduced to master integrals with the help of \texttt{Kira} \cite{Maierhofer:2017gsa,Klappert:2020nbg}
    with \texttt{Fermat} \cite{fermat} on a family-by-family basis. We make sure to reduce 
    to a basis where the dependence on $d$ and the kinematic variable $s$ factorizes utilizing 
    an improved version of \texttt{ImproveMasters}, first developed in Ref.~\cite{Smirnov:2020quc}.
    Afterwards we symmetrize over all families and find 422 master integrals for the non-singlet 
    contribution and 316 for the singlet diagrams.
        In a next step, we set up a systems of differential equations for the master integrals in the variable 
    $\hat{s}=s/m^2$ by calculating the derivatives with the help of \texttt{LiteRed} \cite{Lee:2012cn,Lee:2013mka} 
    and subsequent reduction with \texttt{Kira}. 

Subsequently, the master integrals need to be solved. 
This is achieved using the semi-numerical approach presented in Ref.~\cite{Fael:2021kyg} 
and explained in more detail for the current problem in Ref.~\cite{Fael:2022miw}. 
Let us summarize the main ideas of the approach:
\begin{enumerate}
    \item We calculate boundary conditions for all master integrals at the special 
    point $s=0$. At this special point the master integrals in the non-singlet case 
    reduce to three-loop on-shell propagators, which are well studied in the literature
    \cite{Laporta:1996mq,Melnikov:2000qh,Lee:2010ik}. However, we needed to extend the depth of the $\epsilon$ expansion, since we 
    encountered high spurious poles in the amplitude after reducing to master integrals.
    The results can be found in Ref.~\cite{Fael:2022miw}.
    Since the singlet diagrams have massless cuts, we need to perform an asymptotic expansion 
    around $s=0$ to obtain their boundary conditions.
    \item We calculate symbolic expansions around the point $s=0$ by 
    inserting a suitable ansatz into the system of differential equations.
    By comparing powers in $\epsilon$, the expansion parameter $s$ and possibly 
    logarithms of the expansion parameter, we obtain a system of linear equations 
    for the coefficients of the ansatz. We solve this system of equations with 
    \texttt{Kira} and \texttt{FireFly} \cite{Klappert:2019emp,Klappert:2020aqs}
    in terms of a small set of boundary conditions, which can be determined from step 1.
    \item We calculate symbolic expansions around a new point $s_1$ and match 
    the two expansions numerically at a point where both expansions converge, e.g.~$s_1/2$.
    \item Afterwards, we generate another symbolic expansion at $s_2$ and 
    match it to the expansion around $s_1$ at a point where both expansions converge.
    This way we can map out the whole kinematics of the process.
\end{enumerate}

\begin{figure}[h!]
    \centering
    \includegraphics[width=0.32\textwidth]{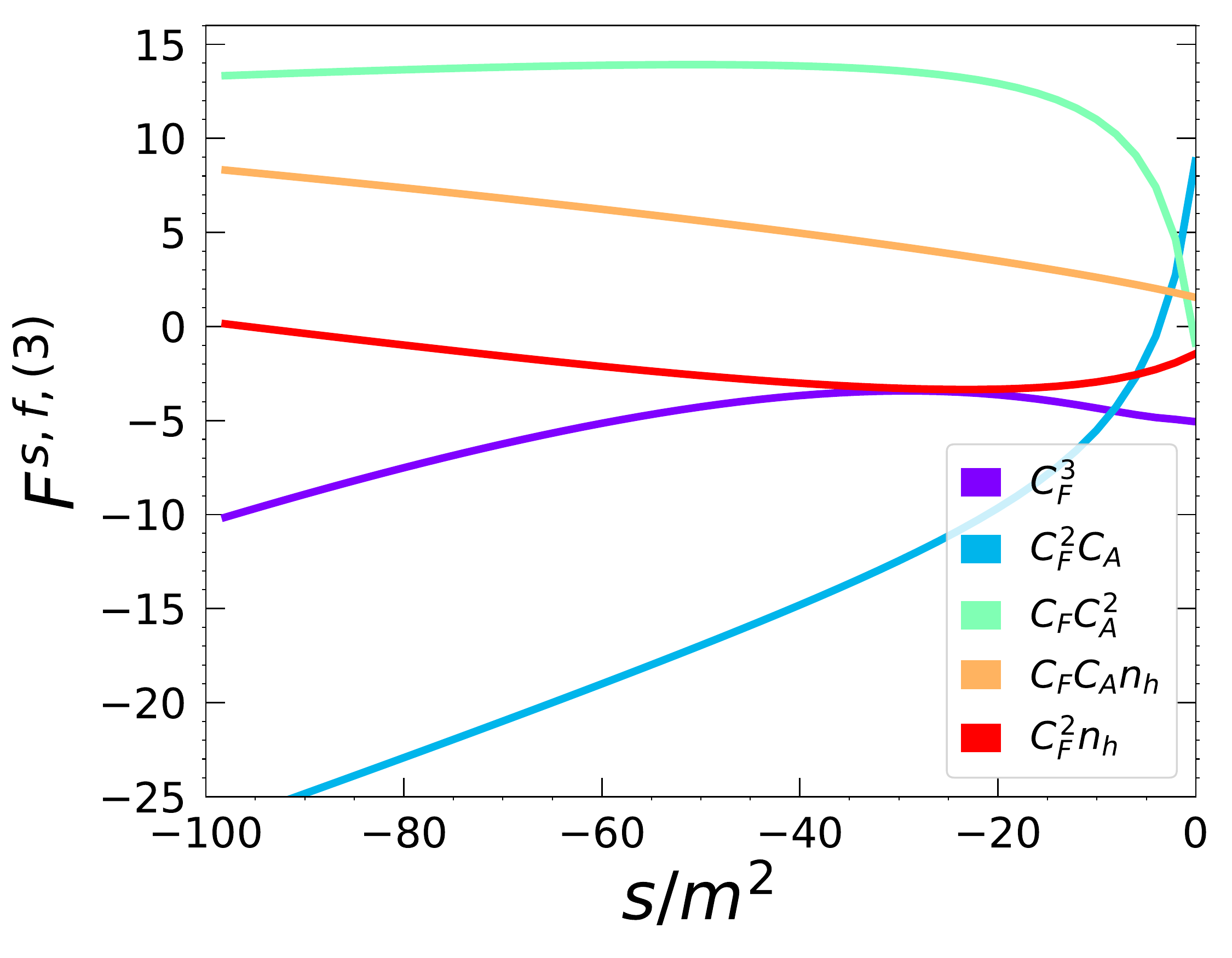}
    \includegraphics[width=0.32\textwidth]{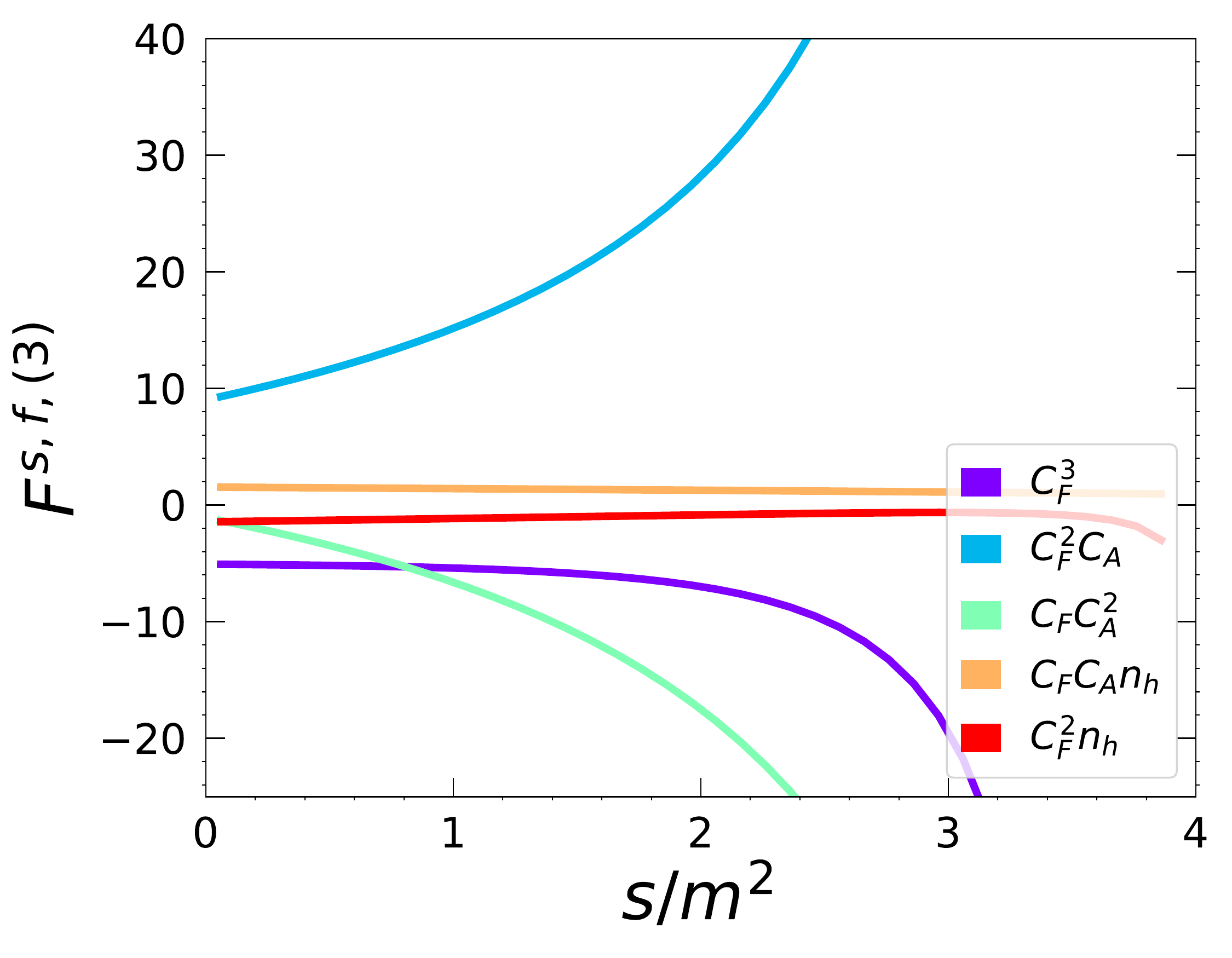}
    \includegraphics[width=0.32\textwidth]{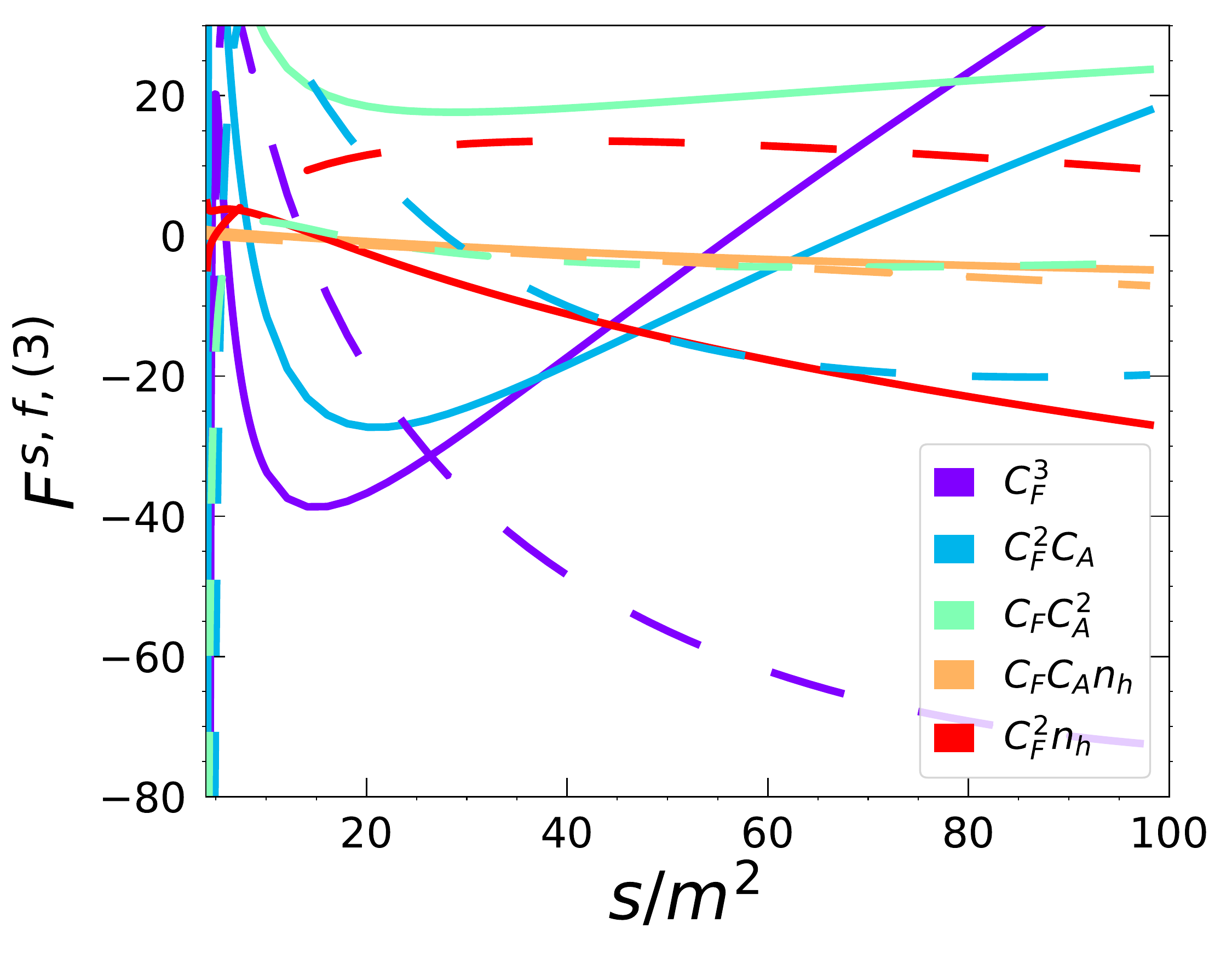}
    \caption{The UV-renormalized and infrared subtracted form factor $F^{s,f,(3)}$. Solid lines correspond to the real part, dashed lines to the imaginary part of the corresponding color factor.
    Taken from Ref.~\cite{Fael:2022miw}.}
    \label{fig::Fs}
\end{figure}

\begin{figure}[h!]
    \includegraphics[width=0.49\textwidth]{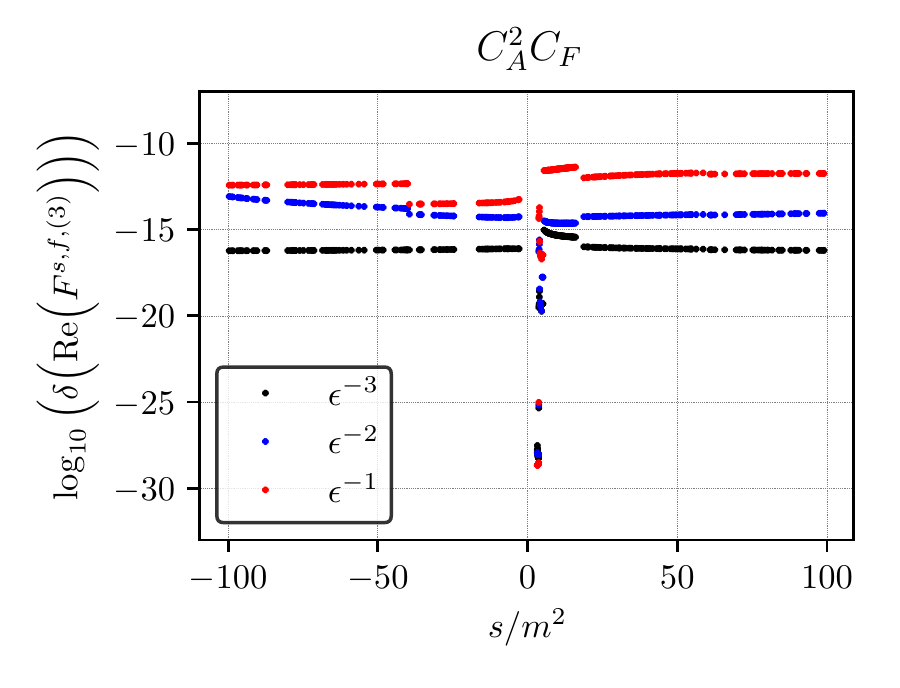}
    \includegraphics[width=0.49\textwidth]{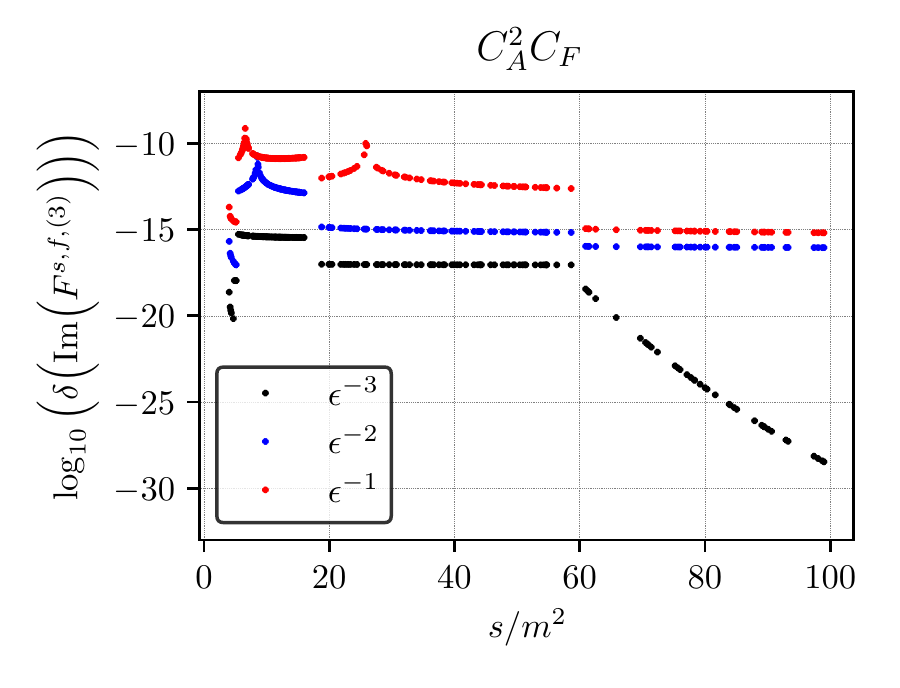}
    \caption{Relative cancellation of the real (left) and imaginary part (right)
    of the poles for the color factor $C_F C_A^2$ of $F^{s,f,(3)}$.
    Taken from Ref.~\cite{Fael:2022miw}.}
    \label{fig::poles}
\end{figure}

\section{Results}
\label{sec:3}
The main results of our method are overlapping series expansions which 
can be used to evaluate the massive form factors at any value of $s$.
In Fig.~\ref{fig::Fs} we show as an example the non-singlet contributions to the 
form factor $F^s$, where the masses and wave function are renormalized on-shell, the current 
in the $\overline{\rm MS}$-scheme and the remaining infrared divergencies
are subtracted by multiplying with a suitably defined $Z$-factor, which can be constructed from 
the cusp anomalous dimension \cite{Polyakov:1980ca,Korchemsky:1987wg,Grozin:2014hna,Grozin:2015kna} 
(see Refs.~\cite{Henn:2016tyf,Fael:2022miw} for the precise definition). 
The resulting finite form factors are labeled with an additional superscript $f$.
Note, that our expansion around $s=0$ is analytic, e.g. for $F^{s,f,(3)}$,
i.e.~the non-singlet contribution to the scalar form factor, we find:
\begin{align}
    &F^{s,f,(3)}\Big|_{s\to0} =
    C_A C_F^2 \Bigg[
      \frac{4 a_4}{3}+\frac{491 \zeta_3}{96}+\frac{19 \pi ^2 \zeta_3}{16}-\frac{45 \zeta_5}{16}+\frac{26117}{4608}-\frac{1193 \pi^2}{576}-\frac{65 \pi ^4}{432}+\frac{l_2^4}{18} \nonumber\\
      &\quad + \frac{31}{36} \pi ^2 l_2^2+\frac{43}{18} \pi ^2 l_2
    \Bigg]
    + C_A^2 C_F \Bigg[
      -\frac{11 a_4}{3}-\frac{947 \zeta_3}{288}-\frac{51 \pi ^2 \zeta_3}{64}+\frac{65 \zeta_5}{32}-\frac{584447}{124416}+ \frac{3011\pi ^2}{3456} \nonumber\\
      &\quad+\frac{179 \pi ^4}{3456}-\frac{11 l_2^4}{72}-\frac{11}{36} \pi ^2 l_2^2+\frac{49}{72} \pi ^2 l_2
    \Bigg]
    + C_F^3 \Bigg[
      12 a_4+\frac{87 \zeta_3}{16}+\frac{\pi ^2 \zeta_3}{16}- \frac{5 \zeta_5}{8}+\frac{55}{96}+\frac{643 \pi ^2}{192} \nonumber\\
      &\quad +\frac{\pi^4}{48}+\frac{l_2^4}{2}-\frac{1}{2} \pi ^2 l_2^2-\frac{15}{2} \pi ^2 l_2
    \Bigg]
    + C_F^2 T_F n_h \Bigg[
      8 a_4+\frac{17 \zeta_3}{18}- \frac{2083}{432}-\frac{52 \pi ^2}{81}-\frac{\pi ^4}{720}+\frac{l_2^4}{3} \nonumber\\
      &\quad -\frac{1}{3} \pi ^2 l_2^2+\frac{8}{9} \pi ^2 l_2
    \Bigg]
    + C_F C_A T_F n_h \Bigg[
      -6 a_4-\frac{199 \zeta_3}{144}+\frac{\pi ^2 \zeta_3}{8} - \frac{5 \zeta_5}{8}+\frac{209857}{15552}-\frac{4351 \pi^2}{1296} \nonumber\\
      &\quad -\frac{\pi ^4}{288}-\frac{l_2^4}{4}+\frac{1}{4} \pi ^2 l_2^2+\frac{32}{9} \pi ^2 l_2
    \Bigg] \nonumber\\
    &\quad + \frac{s}{m^2} \Bigg\{
    C_A C_F^2 \Bigg[
      -\frac{8 a_4}{9}+\frac{2515 \zeta_3}{2304}-\frac{29 \pi ^2 \zeta_3}{144}-\frac{95 \zeta_5}{48}+\frac{11191}{41472}-\frac{15101\pi ^2}{62208}+ \frac{37 \pi ^4}{4320}-\frac{l_2^4}{27} \nonumber\\
      &\quad +\frac{1259 \pi ^2 l_2^2}{2160}+\frac{1409 \pi ^2 l_2}{1440}
    \Bigg]
    + C_A^2 C_F \Bigg[
      \frac{5 a_4}{4}+\frac{8675 \zeta_3}{10368}-\frac{73 \pi ^2 \zeta_3}{1152}+ \frac{125 \zeta_5}{384}-\frac{851465}{279936} \nonumber\\
      &\quad +\frac{130417 \pi ^2}{186624}+\frac{689 \pi ^4}{103680}+\frac{5 l_2^4}{96}-\frac{1}{20} \pi ^2 l_2^2-\frac{12253 \pi ^2 l_2}{8640}
    \Bigg]
    + C_F^3 \Bigg[
      -\frac{29 a_4}{9}-\frac{12401 \zeta_3}{3456} \nonumber\\
      &\quad -\frac{67 \pi ^2 \zeta_3}{288}+\frac{85 \zeta_5}{32}+\frac{22613}{41472}-\frac{69355 \pi ^2}{31104}+\frac{1727 \pi ^4}{25920}- \frac{29 l_2^4}{216}-\frac{1043 \pi ^2 l_2^2}{1080}+\frac{4013 \pi ^2 l_2}{1080}
    \Bigg] \nonumber\\
    &\quad + C_F^2 T_F n_h \Bigg[
      \frac{8 a_4}{9}+\frac{9889 \zeta_3}{6912}-\frac{8059}{5184}-\frac{4261 \pi ^2}{25920} +\frac{7 \pi ^4}{1620}+\frac{\log^4(2)}{27}-\frac{1}{27} \pi ^2 l_2^2+\frac{4}{27} \pi ^2 l_2
    \Bigg] \nonumber\\
    &\quad + C_F C_A T_F n_h \Bigg[
      -\frac{5 a_4}{18}+ \frac{1657 \zeta_3}{1728}+\frac{\pi ^2 \zeta_3}{96}+\frac{5 \zeta_5}{96}+\frac{2257}{3888}-\frac{13663 \pi^2}{25920}-\frac{121 \pi ^4}{51840}-\frac{5 l_2^4}{432} \nonumber\\
      &\quad +\frac{5}{432} \pi ^2 l_2^2+ \frac{55}{108} \pi ^2 l_2
    \Bigg]
    \Bigg\}
    + {\cal O}\left(\frac{s^2}{m^4}\right) + \mbox{$n_l$, $n_l^2$ and $n_h^2$ terms} ,
\end{align}
where $l_2=\log(2)$, $a_4=\mbox{Li}_4(1/2)$ and $\zeta_n$ is Riemann's zeta
function evaluated at $n$ and $C_F=T_F(N_C^2-1)/N_C$, $C_A=2T_F N_C$ are the quadratic Casimir 
operators of the ${\rm SU}(N_C)$ gauge group in the fundamental and adjoint representation, 
respectively, $n_l$ is the number of massless quark flavors, $n_h$ is the number 
of heavy quark flavors with mass $m$ and $T_F = 1/2$. 

There are several checks on our results. 
For example, the coefficient in front of the gauge parameter in the final result 
is smaller than $10^{-18}$ and we can reproduce the known analytic results in the 
planar limit, the contributions $\sim n_l$ and the $n_h^2$ contributions with at 
least 12 digits.
Furthermore, the results are precise enough to calculate the leading and sub-leading
logarithmic corrections in the high energy expansion for the first power suppressed
contributions analytically.
These corrections have been obtained in Refs.~\cite{Liu:2017axv,Liu:2017vkm,Liu:2018czl,Liu:2021chn} 
by considering an involved asymptotic expansion of the Feynman diagrams.
We find agreement except of the quartic mass suppressed corrections to the form 
factor $F_2^{v,(3)}$.
Our results have been confirmed by the authors of Ref.~\cite{Liu:2021chn}. 
More details and analytic expressions for several expansion terms can be found in 
Ref.~\cite{Fael:2022miw}.

The precision of our final results can be estimated from the cancellation 
of the poles in the dimensional regulator $\epsilon$, since they are known 
analytically and have to cancel in the final result.
We use the logarithm to the base 10 of the relative pole cancellation (denoted by $\delta$)  
as a measure of accurate digits. 
A plot of this measure for the form factor $F^{s,f,(3)}$ and the color factor $C_F C_A^2$, 
split into real and imaginary part, can be found in Fig.~\ref{fig::poles}.
We see that the precision for $s < 3 $ and $s > 16$ is highest and 
decreases for the regions between the two thresholds at $s=4m^2$ (two particle threshold)
and $s=16m^2$ (three particle threshold), which are not analytic.
In total we estimate at least 7 significant digits over the whole kinematic range 
for all of our results.
The results of the singlet contributions is significantly higher and estimated to be 
at least 10 digits.

A \texttt{Mathematica} package to evaluate the form factors in the non-singlet and 
singlet case numerically over the full kinematic range of $s$ can be found at: \\
\begin{center}
\vspace{-0.75cm}
\url{https://gitlab.com/formfactors3l/formfactors3l}.
\end{center}

\section{Conclusions and Outlook}
\label{sec:4}
We presented our recent calculation of massive quark form factors at 
$\mathcal{O}(\alpha_s^3)$ which uses a semi-numerical method based on 
series expansions and numerical matching to obtain results for the 
form factors for the whole kinematic range of negative and 
positive values of the virtuality $s$.
We obtain a precision of at least 7 significant digits in the non-singlet 
and 10 digits in the singlet case over the whole kinematic range, respectively. 
However, some kinematic regions are much more precise.
Thus, it is for example possible to extract leading and sub-leading 
logarithmic contributions to the leading and first power suppressed terms
in the high energy expansion analytically, confirming and correcting 
results in the literature.
To complete the calculation of massive quark form factors the singlet 
diagrams where the external current couples to an internal light quark loop
still need to be completed.

\acknowledgments \noindent
I thank Matteo Fael, Fabian Lange and Matthias Steinhauser for the 
enjoyable and productive collaboration. 
The Feynman diagrams have been generated using \texttt{Feyngame} \cite{Harlander:2020cyh}.
This research was supported by the Deutsche Forschungsgemeinschaft (DFG, German ResearchFoundation) 
under grant 396021762 — TRR 257 “Particle Physics Phenomenology after theHiggs Discovery”.

\bibliographystyle{JHEP}
\bibliography{bib}

\end{document}